\PassOptionsToPackage{table}{xcolor}
\documentclass[sigconf]{acmart}

\makeatletter
\def\@ACM@checkaffil{
	\if@ACM@instpresent\else
	\ClassWarningNoLine{\@classname}{No institution present for an affiliation}%
	\fi
	\if@ACM@citypresent\else
	\ClassWarningNoLine{\@classname}{No city present for an affiliation}%
	\fi
	\if@ACM@countrypresent\else
	\ClassWarningNoLine{\@classname}{No country present for an affiliation}%
	\fi
}
\makeatother

\setcopyright{rightsretained}
\copyrightyear{2024}
\acmYear{2024}
\acmDOI{XXXXXXX.XXXXXXX}

\acmISBN{978-1-4503-XXXX-X/18/06}

\makeatletter\def\input@path{{latex/styles/}}\makeatother 
\usepackage{fast-todo}
\usepackage{formatting}
\usepackage{tables}

\usepackage{tikz}
\usepackage{pgf-umlsd} 
\usepackage{hyperref}

\usepackage{cascadia-code}
\usepackage{listings}
\usepackage{color}
\usepackage{textcomp}
\usepackage{lstautogobble} 

\usepackage[T1]{fontenc}

\lstdefinelanguage{json}
{
	morestring=[b]",
	morestring=[d]',
	morecomment=[s]{/*}{*/},
	xleftmargin={0pt},
}

\lstdefinelanguage{JavaScript}{
	morekeywords=[1]{break, continue, delete, else, for, function, if, in,
		new, return, this, typeof, var, void, while, with},
	morekeywords=[2]{false, null, true, boolean, number, undefined,
		Array, Boolean, Date, Math, Number, String, Object},
	morekeywords=[3]{eval, parseInt, parseFloat, escape, unescape},
	sensitive,
	morecomment=[s]{/*}{*/},
	morecomment=[l]//,
	morecomment=[s]{/**}{*/}, 
	morestring=[b]',
	morestring=[b]"
}[keywords, comments, strings]

\lstdefinelanguage[ECMAScript2015]{JavaScript}[]{JavaScript}{
	morekeywords=[1]{await, async, case, catch, class, const, default, do,
		enum, export, extends, finally, from, implements, import, instanceof,
		let, static, super, switch, throw, try},
	morestring=[b]` 
}

\lstalias[]{ES6}[ECMAScript2015]{JavaScript}

\definecolor{mediumgray}{rgb}{0.3, 0.4, 0.4}
\definecolor{mediumblue}{rgb}{0.0, 0.0, 0.8}
\definecolor{forestgreen}{rgb}{0.13, 0.55, 0.13}
\definecolor{darkviolet}{rgb}{0.58, 0.0, 0.83}
\definecolor{royalblue}{rgb}{0.25, 0.41, 0.88}
\definecolor{crimson}{rgb}{0.86, 0.8, 0.24}

\definecolor{backgroundColor}{HTML}{ffffff} 		
\definecolor{identifierColor}{HTML}{1F2328} 		
\definecolor{commentColor}{HTML}{6e7781}			
\definecolor{stringColor}{HTML}{0a3069}				
\definecolor{keywordColor}{HTML}{cf222e}			
\definecolor{valueKeywordColor}{HTML}{0550ae}		
\definecolor{functionKeywordColor}{HTML}{8250df}	
\definecolor{emphasisColor}{HTML}{82071e}			

\lstdefinestyle{base}{
	autogobble,
	showlines=true,
	emptylines=1,
	basicstyle=\scriptsize\ttfamily,
	identifierstyle=\color{identifierColor},
	commentstyle=\color{commentColor}\upshape,
	stringstyle=\color{stringColor},
	keywordstyle=\color{keywordColor},
	keywordstyle=[2]\color{valueKeywordColor},
	keywordstyle=[3]\color{functionKeywordColor},
	emphstyle=\color{emphasisColor},
	extendedchars=true,
	upquote=true,
	tabsize=2,
	showtabs=false,
	showspaces=false,
	showstringspaces=false,
	numbers=left,
	numberstyle=\scriptsize\color{black},
	numbersep=5pt,
	numberblanklines=true,
	captionpos=b,
	xleftmargin=2em,
	xrightmargin=1em,
	frame=single,
	backgroundcolor=\color{backgroundColor},
}

\begin{document}
	
	\title[Passwords Are Meant to Be Secret: A Practical Secure Password Entry Channel for Web Browsers]{Passwords Are Meant to Be Secret:\\A Practical Secure Password Entry Channel for Web Browsers}
	\author{Anuj Gautam}
	\affiliation{\small{University of Tennessee, Knoxville}}
	\author{Tarun Kumar Yadav}
	\affiliation{\small{Brigham Young University}}
	\author{Kent Seamons}
	\affiliation{\small{Brigham Young University}}
	\author{Scott Ruoti}
	\affiliation{\small{University of Tennessee, Knoxville}}
		
	

\begin{abstract}

Password-based authentication faces various security and usability issues. 
Password managers help alleviate some of these issues by enabling users to manage their passwords effectively. 
However, malicious client-side scripts and browser extensions can steal passwords after they have been autofilled by the manager into the web page.
In this paper, we explore what role the password manager can take in preventing the theft of autofilled credentials without requiring a change to user behavior.
To this end, we identify a threat model for password exfiltration and then use this threat model to explore the design space for secure password entry implemented using a password manager.
We identify five potential designs that address this issue, each with varying security and deployability tradeoffs.
Our analysis shows the design that best balances security and usability is for the manager to autofill a fake password and then rely on the browser to replace the fake password with the actual password immediately before the web request is handed over to the operating system to be transmitted over the network.
This removes the ability for malicious client-side scripts or browser extensions to access and exfiltrate the real password.
We implement our design in the Firefox browser and conduct experiments, which show that it successfully thwarts malicious scripts and extensions on 97\% of the Alexa top 1000 websites, while also maintaining the capability to revert to default behavior on the remaining websites, avoiding functionality regressions.
Most importantly, this design is transparent to users, requiring no change to user behavior.
\end{abstract}


	
\maketitle


\section{Introduction}

Despite their many problems~\cite{florencio2007large,dellamico2010password,ur2015added,pearman2017lets}, passwords remain the dominant form of authentication on the Web.
While they may someday be fully replaced by more secure alternatives~\cite{bonneau2012quest}, for now, we must secure the use of passwords.

Password managers seek to improve the security of passwords by improving the usability of password generation, storage, and entry~\cite{oesch2020that,simmons2021systematization}, with the hope that doing so will encourage users to generate random passwords and avoid password reuse.
Additionally, password managers allow users to audit the security of their stored passwords, detecting weak, reused, or compromised passwords.
However, recent research has shown that users avoid using these features, ~\cite{fagan2017investigation,pearman2019why,lyastani2018better,oesch2022observational}, limiting the security benefits of adopting a password manager.

Arising from this state of affairs, we ask the question, \textit{is there anything that password managers can do to increase the security of passwords that does not rely on users to change their behavior?}
In this paper, we start to answer this question by investigating whether password managers can protect the passwords they autofill from theft.
This includes preventing theft from web trackers~\cite{senol2022leaky}, cross-site scripting (XSS) attacks~\cite{help_net_security_2021}, malicious browser extensions~\cite{kapravelos2014hulk,pantelaios2020you}, or compromised JavaScript libraries~\cite{ohm2020backstabber,duan2020towards} (i.e., a supply chain attack).

We start our investigation by exploring the design space for theft-resilient password entry.
We identify five possible designs, evaluating each based on their security and usability.
For security, we consider their ability to prevent theft by phishing, honest-but-curious scripts, malicious scripts (e.g., XSS), and malicious browser extensions.
For deployability, we consider whether these designs require changes to websites, extension APIs, or the browser itself.

Based on our analysis, we identify the following approach as providing the best mixture of security and deployability:
When autofilling a password, the password manager instead autofills a fake password.
It then provides the browser with (a) the fake password, (b) the real password, and (c) the origin bound to the password.
The browser will then examine outgoing web requests for that web page, looking for the fake password in the web request body.
If found, it replaces the fake password with the real password if and only if the password is being sent to the appropriate origin (as specified by the password manager).
Critically, at no point is the real password ever contained in the DOM---preventing theft by client-side scripts---or within the web request bodies as seen by \texttt{webRequest} API---preventing theft by extensions.\footnote{This design is inspired by the work of Stock and Johns~\cite{stock2014protecting}. A comparison to their work is described later in this paper (\S\ref{sec:comparison}).}
Additionally, this design does not require modifying user behavior or websites.

To demonstrate the feasibility of this design, we forked and modified the Firefox browser to implement the above functionality.
We also modified BitWarden, an open-source password manager, to work with our modified Firefox implementation.
Importantly, our proof-of-concept implementation does not require modifying user behavior or websites.

We empirically evaluated the security of our design, demonstrating that it stops password exfiltration, both by DOM- and extension-based adversaries.
Moreover, we describe how to prevent attacks by adversaries who are aware of this defense and may try to subvert it.
We also evaluate our tool on the 573 sites with login forms from the Alexa Top 1000, showing that it is compatible with 97\% of those sites.
For the remaining 3\% of websites, it is easy for the password manager to revert to the existing behavior, preventing any functionality regression.

We conclude the paper by discussing lessons learned from our design exploration and implementations.
We also describe potential avenues for future research that, similar to our work, could modify the browser to add first-class support for authentication.

In summary, the contributions of our paper are as follows:

\begin{enumerate}
	\item \textbf{Threat model identification and design space exploration.}
	We identify a threat model for theft-resilient password entry.
	Using this model we identify five different approaches, evaluating each based on security and deployability.
	Several of these designs are inspired by work from Stock and Johns~\cite{stock2014protecting}, though we expand on this work, demonstrating limitations in the original proposal and then showing how those limitations can be addressed in different ways to create three of the five designs.
	
	\item \textbf{Proof-of-concept implementation.}
	We create a proof-of-concept implementation of the design we believe has the best combination of security and deployability.
	This includes both a modified version of the Firefox browser and the BitWarden password manager.
	While the final code diff is rather straightforward, creating it was not.
	Doing so required hundreds of hours of engineering effort and many discussions with Firefox developers, many of whom initially believed that what we were proposing would require too much of changes all across the codebase.
	
	\item \textbf{Real-world evaluation.}
	We conducted empirical evaluations to demonstrate that our tool did not interrupt normal authentication flows, including authentication to websites we created and 554 out of 573 websites pulled from the Alexa top 1000 list.
	Additionally, we implement proof-of-concept password-theft attacks for malicious client-side scripts and browser extensions, demonstrating that these attacks worked without our proof-of-concept implementation but were stopped by our implementation.
	
\end{enumerate}

All our source code is publicly available at [redacted].

\section{Background and Related Work}
First, we describe the password entry workflow that is being secured in our work.
Next, we describe background and related work on browser-based password exfiltration and password managers.
We then conclude with background on how browsers function, including form submission, the \texttt{webRequest} API, and browser extension permissions.

\subsection{Password Entry Workflow}
The password entry workflow can be split into the following steps:

\begin{enumerate}
	\item The user visits a web page that has a \texttt{form} with an \texttt{input} element to enter their password.
	\item The user enters their password by (a) manually typing it, (b) autofilling it from their password manager, or (c) copying and pasting it from their password manager. At this point, the password is stored within the web page's DOM.
	\item The user submits the form. This causes the browser to process the form and send a web request to the server that contains the entered password.
	\item The password is transferred over the network (preferably using a TLS connection).
	\item The server receives and processes the password as it deems appropriate.
\end{enumerate}

Note this flow also applies to account creation, which has the same general process for entering and transmitting passwords.

\subsection{Browser-Based Password Exfiltration}

Examining the literature, we identify three avenues by which passwords can be exfiltrated from the browser.
This does not include threats outside of the browser such as phishing~\cite{akerlof2015phishing}, man-in-the-middle network attacks~\cite{oneill2016tls}, or compromised execution environments~\cite{fahl2013hey,mysk2020clipboard}.
While each threat has extensive related work, a high-level understanding of them (as held by most readers) is sufficient to understand the remainder of this paper.
As such, we omit discussing them in this section for brevity.

\subsubsection{Web Trackers}
Web trackers are scripts used to track users across different websites, primarily to more effectively serving advertisements.
Trackers collect various information and interactions of the user without the user even knowing about it.
Senol et al.~\cite{senol2022leaky} analyzed form submissions on 100,000 popular websites, investigating whether web trackers on those pages would exfiltrate user passwords.
They find that at least 2--3\% of those pages include web trackers that exfiltrate passwords.
While the motivations for this exfiltration are largely unknown, in many cases, it likely represents an honest-but-curious attacker model---i.e., web trackers are simply grabbing whatever information they can and are not targeting passwords specifically.
Regardless, as shown by Dambra et al.~\cite{dambra2022sally}, users encounter these web trackers frequently, indicating a need to protect user passwords against these honest-but-curious trackers.

\subsubsection{Malicious Client-Side Scripts}
Malicious scripts running within a web page can steal user passwords by extracting them from the document object model (DOM) after they have been typed into a form by the user or autofilled by a password manager.
There are several sources of malicious client-side scripts.

The most common source of malicious client-side scripts is cross-site Scripting (XSS) attacks.
In these attacks, an adversary coerces a website to include attacker-controlled scripts in a web page's DOM, whether by tricking users into clicking a link with the malicious script (reflected XSS attack) or uploading the malicious script to the website (stored XSS attack).
While defenses for XSS attacks are well known, the OWASP foundation consistently ranks them in its top 10 web application security risks~\cite{xssowasp}, and hundreds of XSS attacks have been reported in January of 2024 alone~\cite{CVEdetails2024}. 

Another common source of malicious client-side scripts are websites that use third-party libraries, with such libraries being ubiquitous~\cite{lemos2021dependency}.
While libraries produced by an adversary are transparently dangerous, more concerning are supply chain attacks~\cite{ohm2020backstabber,duan2020towards}.
In these attacks, an adversary will compromise an otherwise benevolent software library; then, when websites relying on this library are updated, they will also become compromised.
These attacks are already somewhat common~\cite{ohm2020backstabber,duan2020towards}, with WhiteSource~\cite{whitesource2022vulnerablenpm} identifying 1300 malicious JavaScript libraries in 2021.

While supply chain attacks are a problem for client-side and server-side libraries, in this paper we are primarily concerned with client-side supply chain attacks.
Web client-side supply chain attacks are especially concerning as it is a common practice to load libraries from external sources.\footnote{For example, see usage instructions for Bootstrap at \url{https://getbootstrap.com/}.}
If a website adopts this practice, the website will be immediately compromised as soon as the library is compromised, without needing to wait for the website to explicitly update to the compromised version of the library.

\subsubsection{Malicious Browser Extensions}
The final avenue for password theft is malicious browser extensions.
Browser extensions have two avenues for exfiltrating user passwords.
First, they can inject client-side scripts into web pages, stealing passwords in the same way as other malicious client-side scripts.
Alternatively, they can inspect the body of outgoing web requests, stealing passwords found in those bodies.

There are many instances of malicious browser extensions used by millions of users on official Chrome/Firefox extension stores.
In 2020, 500 Chrome browser extensions were discovered secretly uploading private browsing data to attacker-controlled servers and redirecting victims to malware-laced websites~\cite{ositcom}.
Also, Awake has identified 111 malicious Chrome extensions that take screenshots, read the clipboard, harvest password tokens stored in cookies or parameters, grab user keystrokes (like passwords), etc.~\cite{awakesecurity}. 
These extensions have been downloaded over 32 million times.
Finally, in 2021, Cato's analysis of network data showed that 87 out of 551 unique Chrome extensions used on customer networks were malicious~\cite{catonetworks}.

In addition to inherently malicious extensions, extensions can also be compromised through supply chain attacks.

\subsubsection{Relation to Our Work}
The above research demonstrates that there is a critical need to protect passwords from being exfiltrated by honest-but-curious or malicious entities.
In this work, we demonstrate how password managers can be modified to prevent password exfiltration for web trackers, malicious content scripts, and malicious browser extensions.

\subsection{Password Managers}

Password managers serve to help users (a) create random, unique passwords, (b) store the user's passwords, and (c) fill in those passwords.
On desktops, password managers are implemented as browser extensions.
The browser does not provide any password management APIs for these extensions to use~\cite{oesch2020that}.
On mobile, there is first-party support for password managers, though this support has significant security issues~\cite{oesch2021emperors}.

\subsubsection{Security}
Password managers have the potential to provide strong security benefits, but also have the potential to act as a single point of failure for users' accounts~\cite{stock2014protecting,li2014emperor,silver2014password,oesch2020that,oesch2021emperors}.
In particular, when passwords are autofilled into websites they are vulnerable to theft by JavaScript and extensions.
In the most alarming case, if the password manager fails to require user interaction before autofilling passwords and allows passwords to be filled into iframes, it opens users to password harvesting attacks that can surreptitiously steal many if not all their passwords~\cite{stock2014protecting,oesch2020that}.
This problem can be made even worse when the operating system enforces incorrect behavior, such as in mobile devices~\cite{oesch2021emperors}.

\subsubsection{Usability}
Simmons et al.~\cite{simmons2021systematization} systematized password manager use cases.
They found that today's managers poorly supported many password manager use cases and that even when supported, they were often targeted at experts rather than the lay users the tools claimed to support.


Lyastani et al.~\cite{lyastani2018better} instrumented a password manager to collect telemetry data regarding password manager usage.
Their results show that users underutilized password generation.
This phenomenon was partly explained in research by Oesch et al.~\cite{oesch2022observational} that surveyed users and showed that many users avoid the security-critical functionality of password managers, such as password generation or password audits because they felt these features were too difficult to use.
Instead, they focused on features with the highest usability, such as autofill.


\subsubsection{Relation to Our Work}
Research into the security of password managers over the last decade has consistently shown that the autofill process is a key component of security issues with password managers~\cite{stock2014protecting,li2014emperor,silver2014password,oesch2020that,oesch2021emperors}.
In this paper, we explore how the autofill process can be transformed from a weakness of password managers to one of their strengths, providing benefits not available for manual entry.
Critically, this security benefit is available without any change to user behavior, which is not the case for other password manager security benefits~\cite{simmons2021systematization,oesch2022observational}.

\subsection{Relation to Stock and John's Work}\label{sec:comparison}
Our desire to investigate how password managers could be used to strengthen password entry (as opposed to generation and storage) was motivated by the excellent work of Stock and Johns~\cite{stock2014protecting}.
In their paper, Stock and Johns propose having the password manager inject a random value in place of the password, that will only be replaced with the real password during network transmission.
Design \#4, as described in \S\ref{sec:design} is based directly on Stock and Johns' proposal, with only a few minor tweaks.
As such, it is natural to ask what scientific contributions this paper makes compared to that paper.
Below we summarize the key ways we extend Stock and Johns' work:

\begin{itemize}
	\item We identify and describe the threat model for securing autofilled passwords (\S\ref{sec:threat}).
	
	\item We perform a design space search for solutions to securing autofilled passwords (\S\ref{sec:design}). In this process, we include Stock and John's proposal, two other approaches we create inspired by Stock and Johns' proposal, and two unrelated approaches. We evaluate and compare each design's strengths and weaknesses, demonstrating why our proposed Design \#5 is more secure and functional than other designs, including that of Stock and Johns.
	
	\item Our Design \#5 (\S\ref{sec:design5}) protects against malicious extensions, something not possible for the design proposed by Stock and Johns. We also evaluate browser extensions found in Chrome Web Store to demonstrate the feasibility of this threat (\S\ref{sec:permissions}). Finally, we empirically demonstrate that our implementation protects against this threat (\S\ref{sec:security-evaluation}).
	
	\item Our security evaluation considers how attackers would try to circumvent the proposed design using a reflection attack and discuss how password managers could mitigate these risks (\S\ref{sec:reflection}). Stock and Johns' work did not consider how an adversary aware of the defense could circumvent it.
	
	\item Our solution is functional in modern browsers (\S\ref{sec:functional-evaluation}). In contrast, Stock and Johns' proposed solution relies on functionality intentionally removed by browser makers. While our paper might make it seem like the idea to move this functionality deeper into the browser is obvious, the lack of any such proposals suggests that it may not be so. Moreover, as we found in our efforts, modifying the browser is challenging for researchers and practitioners with many potential pitfalls. Thus our implementation, including the code we created, is beneficial both for other researchers to build upon and for browser makers to adopt.
\end{itemize}

\subsection{Browser Background}
Below, we describe how the browser handles the webRequest API and extension permissions.
These are each important parts of our design space exploration and proof-of-concept implementations.

%

\subsubsection{\texttt{webRequest} API}
\label{sec:webrequest}

\begin{figure}
	\centering
	\includegraphics[width=6.5cm]{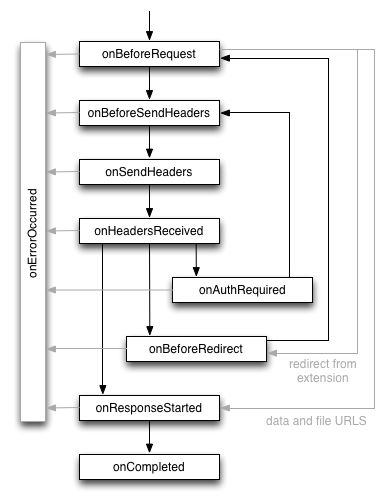}
	\caption{Web Request API flow~\cite{chrome_webrequest}}
	\label{fig:webrequest}
\end{figure}

The \texttt{webRequest} API is a browser extension-only API that lets extensions read, modify, or cancel web requests and responses.
To access this API, extensions register event listeners for one or more stages of the web request processing lifecycle (see Figure~\ref{fig:webrequest}).

In this lifecycle, the \texttt{onBeforeRequest}, \texttt{onBeforeSendHeaders}, and \texttt{onSendHeaders} events happen before the request is sent and provide access to the request body, allowing passwords included there to be exfiltrated.
The \texttt{onBeforeRequest} and \texttt{onBeforeSendHeaders} allow extensions to redirect or cancel the request, while the latter also allows modification of request headers.
The remaining stages are focused on what happens after the request has been submitted and lacking access to the request body is less important in the context of secure password entry.

Critically, none of these stages allow the request body to be modified.
By the time \texttt{onBeforeRequest} is reached, the form submission process has already created the \texttt{FormSubmission} object, and the extensions are only provided with a read-only copy of the data in this object (not the object itself).
The reason behind not allowing the request body to be changed is unknown, but requests to add this functionality have gone unfilled over the last decade and there is little evidence that it will be added~\cite{chromium91191webrequest, mozilla1376155webrequest}.
Interestingly, functionality allowing the modification of request bodies existed in older browser versions (e.g., Firefox, Safari, Internet Explorer) but was removed when these browsers migrated the Chromium model for browser extensions.

While the request body cannot be modified, the response body can be modified in the \texttt{onResponseStarted} stage.
This allows extensions to arbitrarily inject client-side scripts into the web page.
However, this ability to modify response bodies requires additional permissions (\texttt{declarativeNetRequest}) on top of those required to read the body.


\subsubsection{Extension Permissions}\label{sec:permissions}

There are a couple of limitations on what a malicious browser extension can do.
Most importantly, all extensions are given a unique origin, preventing one extension from accessing the data or scripts of another extension.
As such, the browser's same origin policy prevents malicious extensions from directly accessing each others' data or scripts.
The only way for a malicious extension to gain access to the passwords stored by a password manager is for that manager to copy those passwords to an origin where the malicious extension has access (i.e., the webpage).

The second limitation is that for a malicious extension to exfiltrate passwords, they need to request certain permissions in their manifest file~\cite{chromeManifest}:

\begin{itemize}
	\item The \texttt{scripting} permission allows the injection of client-side scripts on any web page~\cite{chromeScriptingAPI}.
	\item The \texttt{activeTab} permission allows the injection of client-side scripts on the root web page of the actively focused tab~\cite{chromeActiveTab}. 
	\item The \texttt{content script} attribute injects a specified client-side script into webpages with a matching origin (wildcards are allowed)~\cite{chromeContentScripts}.
	\item The \texttt{declarativeNetRequest} permission allows the modification of web response bodies, which can be used to inject client-side scripts~\cite{chromeDelaritiveNetRequestAPI}.
	\item The \texttt{webRequest} permission allows reading web request bodies (which can contain passwords)~\cite{chromeWebRequestAPI}.
\end{itemize}

For an extension to be granted the permissions listed in its manifest, the user will need to approve these permissions during installation or when they are first used.
However, research has shown that users struggle to understand these permissions and are likely to grant them without much consideration~\cite{korir2022empirical}.
Still, permissions are useful.
If an extension is compromised through a supply chain attack, the attacker will not be able to change the manifest and will be constrained by the extension's existing permissions.

To understand the prevalence of these permissions in existing extensions, between August 28, 2022, to September 13, 2022, we collected 101,414 browser extensions from the Chrome webstore.
We then extracted and analyzed their manifest files.
We also estimate user counts based on data fetched from CRXcavator\footnote{\url{https://crxcavator.io/}} (pulled May 03, 2023).

We identified 12,576 extensions that can inject client-side scripts on any web page, including 55 with at least 500,000 users.
We also identified 4,169 extensions that can read any web request's body, including 19 with at least 500,000 users.
Of these extensions, 1,410 can read any web request's body but not inject client-side scripts, including 6 with at least 500,000 users.

\section{Threat Model}
\label{sec:threat}

Our threat model focuses on password managers and the exfiltration of users' passwords during the password entry workflow.
This exfiltration can occur (i) after the password has been entered and is stored in the web page's DOM, (ii) as the authentication web request is processed in the browser, (iii) when the password is transmitted over the network, and (iv) when the password is received by the server.
Since we focus on securing password entry for password managers, we do not consider the case where the user manually types the password or how the server handles the password after the server receives it (e.g., whether they store passwords salted and hashed).

In our threat model, we are concerned with the following adversaries:

\begin{enumerate}
	\item \textbf{Honest-but-curious entities} (e.g., web trackers~\cite{senol2022leaky}).
	This entity can read the contents of any DOM element, including the password after it has been entered into the web page.
	Unlike the other adversaries, as an honest party, this adversary will not circumvent defensive measures designed to protect the password.
	
	\item \textbf{DOM attacker.}
	This adversary has full control of the web page's DOM, able to read and modify the DOM at will.
	They can steal the password if it is ever included in the web page's DOM.
	They can also attempt to trick password managers into autofilling passwords outside the legitimate login page~\cite{silver2014password,stock2014protecting,oesch2020that}.
	However, this adversary's capabilities are limited by security primitives built into the browser---they cannot violate the same origin policy (cannot directly access the password manager's data) and do not have access to web requests or responses that the attacker did not generate.
	
	\item \textbf{Extension attacker.}
	This attacker can read the headers and body of all web requests and responses, allowing them to steal any passwords in web request bodies (as visible through the \texttt{webRequest} API).
	While a malicious extension can also inject a client-side script, this capability is already covered by the DOM attacker, so when talking about the extension attacker, we are focused on their ability to read web requests and responses.
	
	\item \textbf{Man-in-the-middle (MitM) attacker.}
	This is a network attacker who sits between the user's device and the server.
	They can eavesdrop and modify network traffic as it is transmitted between the two.
	As we evaluate the security of password entry defenses, we consider two variants of this adversary: one that cannot break TLS-encrypted communication and one that can (e.g., an attacker who leverages a substitute certificate attack~\cite{oneill2016tls}).
	
	\item \textbf{Phisher.}
	This adversary has full control of the web page visited by the user and the server where the password will be sent, being able to steal the password if it is entered into the web page or transmitted to the server.
	While this attacker must convince a user to visit the phishing web page, we assume this is feasible~\cite{akerlof2015phishing}.
	For simplicity, we combine \textbf{compromised websites} with phishers as the two attackers have the same capabilities.
\end{enumerate}

In addition to these adversaries, we are also concerned with a \textbf{supply chain attacker}.
This attacker compromises a library used by a server, a client-side web page, or a browser extension.
At this point, the attacker has the same capabilities as a phisher, DOM attacker, and MitM attacker, respectively.
Thus, we don't evaluate this adversary separately from those adversaries.
However, we mention them here to emphasize the feasibility of the other adversaries, as supply chain attacks can put users' passwords at risk, even if the user exercises extreme caution (e.g., vetting all links, disabling client-side scripts, or vetting all extensions).

In our threat model, we intentionally consider a wide range of adversaries to better evaluate the design space for password autofill defenses.
However, we consider several threats as out of scope.
First, we consider compromised browsers and operating systems as out of scope.
If these items are compromised, there is no need to steal credentials during the autofill process, as they can simply be stolen from memory after the password manager decrypts them.

Second, we do not consider session hijacking attacks.
Poor cookie hygiene (e.g., failing to use \texttt{HTTPOnly} cookies) can allow session theft by any of our identified adversaries.
However, existing defenses can prevent this attack for any adversary, up to and including a malicious extension (through token-bound cookies~\cite{RFC8471}).

Even if we exclude malicious extensions that can exfiltrate session cookies (using the \texttt{cookie} permission), there are still 11,452 extensions that can inject client-side scripts, 3,568 that can read requests' bodies, and 1,294 that can read web requests' bodies, but not inject client-side scripts.
This indicates that even with this carveout there are a substantial number of extensions that satisfy our threat model.

Finally, we note that while malicious extensions may appear similar to a compromised browser, this is far from the case.
First, most extensions have limited permissions, limiting the damage they can do when compromised.
Second, even after granting a malicious extension every possible permission, it is still sandboxed by the same origin policy, meaning it cannot steal data stored by other extensions (i.e., the password manager).
For these reasons, we believe it is reasonable to consider malicious extensions in scope but compromised browsers out of scope.


\section{Design Space Exploration}
\label{sec:design}

\begin{table}
	\setupratingstable
	\adjustbox{max width=\columnwidth}{
		\begin{tabular}{ll|lllll|ll|}
			\cline{3-9}
			& & \multicolumn{5}{c|}{Security} & \multicolumn{2}{c|}{Deploy}
			\\ \cline{3-9}
			& Design	 					
			& \rowheader{Protection from honest-but-curious entity}
			& \rowheader{Protection from DOM attacker}
			& \rowheader{Protection from extension attacker}
			& \rowheader{Protection from MitM attacker}
			& \rowheader{Protection from phisher}
			
			& \rowheader{No changes to websites}
			& \rowheader{No changes to browsers}
			\\ \hline
			
			1. & Zero-knowledge proof
			& \sfull	& \sfull	& \sfull	& \sfull	& \sfull	
			& \snone	& \sfull
			\\ \hline
			
			2. & Modified form handling			
			& \sfull	& \snone	& \snone	& \snone	& \spart
			& \sfull	& \spart
			\\ \hline
			
			3. & JS-based nonce injection		
			& \spart	& \spart	& \snone	& \spart	& \spart
			& \sfull	& \sfull
			\\	
			
			4. & API-based nonce injection		
			& \sfull	& \sfull	& \snone	& \spart	& \sfull
			& \sfull	& \spart
			\\		
			
			5. & Browser-based nonce injection	
			& \sfull	& \sfull	& \sfull	& \spart	& \sfull
			& \sfull	& \snone
			\\ \hline \hline
			
			& Current password manager autofill
			& \snone	& \snone	& \snone	& \snone	& \spart
			& \sfull	& \sfull
			\\
			
			& Two-factor authentication (2FA)	
			& \snone	& \snone	& \snone	& \snone	& \snone
			& \snone	& \snone
			\\
			
			& Phishing-resistant 2FA			
			& \spart	& \spart	& \spart	& \spart	& \spart
			& \snone	& \snone
			\\ \hline
			
		\end{tabular}	
	}
	
	\begin{tabular}{ll}
		\sfull & Fully achieves the property \\
		\spart & Achieves the property with some limitations \\
		\snone & Fails to achieve the property \\
	\end{tabular}
	
	\caption{An evaluation of the five designs based on security and deployment. Also includes an evaluation of 2FA as a comparison point.}
	\label{tab:properties}
\end{table}

Based on our threat model, we explored the design space for implementing password entry defenses in password managers.
A review of the password manager and password exfiltration literature guided this exploration.
Repeated discussions within our research group also informed it.
In total, we identify five high-level designs for securing password autofill.

We then evaluate these designs along two axes: security and deployability.
For security, we consider whether these designs could survive attacks by the attackers identified in our threat model.
For deployability, we consider whether these designs avoid changes to websites and the browser.
Avoiding changes to websites is critical, as requiring all websites to adopt new technology is unlikely to succeed, as can be seen with the lack of adoption for many proposed authentication technologies~\cite{bonneau2012quest}.
Avoiding changes to the browser is also ideal, though it is easier to change browsers than all websites.
Moreover, we also distinguish between changes to the browser that affect the UI processes (similar to user-space processes) and those that affect the core browser process (similar to the kernel process).

A summary of our evaluation is shown in Table~\ref{tab:properties}.
At the bottom of this table, we compare the identified designs against existing approaches for security authentication.
For password manager autofill, we note that there is partial protection from phishing attacks as the manager should not autofill passwords on phishing websites~\cite{oesch2020that,oesch2021emperors}. However, the user can still copy and paste passwords into password into the phishing website.
We also compare against 2FA (discussed at the end of this section) to highlight that moving to 2FA does not solve this issue as some might assume.

\subsection{Design \#1: Zero-Knowledge Proof}
The most secure approach for password authentication is using zero-knowledge proofs (ZKPs), where the user does not reveal their password to the server.
Examples of this approach are augmented password-authenticated key exchange (PAKE) protocols~\cite{wu1998secure,jarecki2018opaque}.

In this design, websites add an endpoint supporting authentication using a ZKP.
This endpoint is not a web page, but rather a method by which the password manager could directly authenticate on the user's behalf using the password stored in the manager.
Once authentication succeeds, the password manager sets a session cookie for the domain to create an authenticated session.
Notably, the password is never placed in the DOM.

\paragraph{Security evaluation.}
As the password is never present in the DOM, it cannot be stolen by attackers that rely on DOM-based exfiltration (honest-but-curious entity, DOM attacker).
Similarly, the zero-knowledge nature of the authentication protocol means that no information sent over the network can be used to derive the password, preventing this avenue of exfiltration (extension attacker, MitM attacker).
Finally, even if a ZKP is performed with a phisher, they will gain no knowledge of the password, protecting against phishing attacks.

\paragraph{Deployability evaluation.}
This design requires all websites to add an endpoint for conducting authentication using a ZKP, which is likely a non-starter.
Note, PAKE-based ZKP for passwords have existed for decades but have never seen widespread adoption~\cite{hao2022sok}.
On a more positive note, this design could be implemented with existing browser functionality.

\subsection{Design \#2: No-Script Form Attribute}
To stop DOM-based exfiltration of passwords, the browser could add support for a new attribute on forms or input elements (e.g., \texttt{noscript}) that prevents scripts from accessing the values stored in that form or input element.
The password manager would set this attribute on password fields before autofilling the password.
At a high level, this approach is similar to the shadow DOM~\cite{shadowDOM} that shows one view of the DOM to the user and another to scripts on the page.
From the user's, web page's, and password manager's perspective, everything would work as it always has, except with an extra layer of password protection.

\paragraph{Security evaluation.}
This approach would protect against an honest-but-curious entity, as they would not have access to the protected value and would not try to circumvent this defense.
However, this approach provides little to no security for the other attacks.
First, a DOM attacker could simply create look-alike forms, display those over the actual form (either by removing it or using a higher z-index), and then steal passwords from the unprotected form.
Second, the password is still sent over the wire allowing access by the extension attacker, the MitM attacker, and the phisher (assuming the user copies and pastes the password).

\paragraph{Deployability evaluation.}
Adding a form attribute will require modifying the browser's DOM code (UI process) but not code that executes in the browser's core process.
As the manager is responsible for updating forms with this new attribute to form elements, no changes are needed to websites.
However, if the web page relies on scripts to access and submit the authentication request, as opposed to letting the form do so itself, the web page's functionality could break.

\subsection{Design \#3--5: Nonce Injection}

Nearly a decade ago, Stock and Johns~\cite{stock2014protecting} evaluated the security of password manager autofill, finding that managers autofilled passwords into malicious websites under many conditions.
At the end of their paper, they proposed a possible solution to this problem:

\begin{enumerate}
	\item When an autofill operation is triggered, a random placeholder for the password (\emph{a password nonce}) is generated.
	\item The password nonce is autofilled into the web page.
	\item The manager will scan outgoing web requests looking for the password nonce.
	\item If the password nonce is detected while being transmitted to an origin that matches the origin for the real password, the manager will replace the nonce with the password.
\end{enumerate}

While the implementation approach proposed by Stock and Johns is not possible (browser extensions cannot modify web request bodies), the core idea remains sound.
Inspired by Stock and Johns' proposal, we identified three designs that autofill password nonces to secure password entry.
Each of these approaches has different security and deployability trade-offs, with Stock and Johns' proposal aligning with Design \#4.

\subsubsection{Design \#3: JavaScript-Based Nonce Injection}
In this design, the password manager will inject a script into the web page when autofilling the password nonce into the web page.
This script has the following responsibilities:

\begin{enumerate}
	\item If the password form already has an \texttt{onsubmit} method, the script will store this method.
	\item Store any existing submit event listeners, then remove all these event listeners from the form.
	\item The script will set the \texttt{onsubmit} method for the form. The new \texttt{onsubmit()} method will perform the following operations:
	\begin{enumerate}
		\item Call the stored \texttt{onsubmit} method, if any.
		\item Call any stored submit event listeners.
		\item Replace the password nonce with the actual password as long as the form data will be submitted to an appropriate origin.
	\end{enumerate}
	\item Modify the DOM so that it is not possible to replace the new \texttt{onsubmit} method. If another script attempts to do so, the \texttt{onsubmit} method stored by this script will be replaced instead.
	\item Modify the DOM so that event listeners can neither be added nor removed from the form, with attempts to do so simply updating the list of event listeners stored by this script.
\end{enumerate}

Items 1--2 and 4--5 are necessary to ensure that no other scripts run after the replacement of the actual password occurs.

\paragraph{Security evaluation.}
This design protects against DOM-based exfiltration in so much as it can ensure that no other scripts can access the DOM after the submission occurs.
While we describe how this could be done, there is no guarantee that this approach will always work.
For example, the browser could change the form submission process, adding new avenues for scripts to run after the password replacement occurs.
In this case, the password might even leak to an honest-but-curious entity.
Alternatively, malicious scripts could look for ways to prevent the password manager's script from preventing modifications to the \texttt{onsubmit} method or event listeners (which run after \texttt{onsubmit} by default). 
While the manager could then update the script it injects to handle these, this sort of cat-and-mouse situation is never ideal.
As such, we rate this design as providing limited protection against honest-but-curious entities and DOM-based attackers.

\begin{figure}
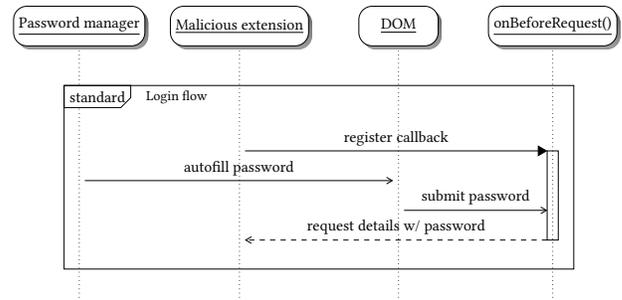

	\centering
	\adjustbox{max width=\columnwidth}{
		\begin{sequencediagram}
			\tikzset{inststyle/.append style={
					drop shadow={top color=gray, bottom color=white}, 
					rounded corners=2.0ex
				}
			}   
			
			\newinst{p}{Password manager}
			\newinst[0.5]{m}{Malicious extension}
			\newinst[1]{d}{DOM}
			
			\newinst[1]{b}{onBeforeRequest()}
			
			\begin {sdblock}{ standard }{Login flow}
			\begin{call}{m}{register callback}{b}{request details w/ password}
				\mess{p}{autofill password}{d}
				\mess{d}{submit password}{b}
			\end{call}
		\end{sdblock}
	\end{sequencediagram}
}

\caption{Diagram illustrating how an attacker can use an \texttt{onBeforeRequest} callback to exfiltrate passwords.}
\label{fig:exfiltrate}
\end{figure}

The real password is still sent in a web request, allowing exfiltration by an extension attacker (see Figure~\ref{fig:exfiltrate}).
Additionally, a MitM attacker can also steal the password during transmission.
While password managers could only replace passwords if they are going to be sent over a valid TLS connection, the connection is often not properly implemented~\cite{oesch2020that} and would not protect against a MitM attacker that can attack TLS connections.

Against a phisher, this design has the same properties as existing password entry using a password manager---the password will not be autofilled on a phishing website. However, if the user copies and pastes it, the phisher will still get the password.
It is not possible to copy a password nonce, as the manager will have no way of knowing on which form to register the replacement of that nonce (though it could try to guess with varying success).

\paragraph{Deployability evaluation.}
This approach requires no modification of anything but the password manager.
From this perspective, it is the most deployable of all the designs we identified.
Still, like Design \#2, this approach could break websites that rely on scripts to submit the password instead of the form.

\subsubsection{Design \#4: API-Based Nonce Injection}
Design \#3 can be improved by changing where nonce replacement happens from the beginning of the form submission process (i.e., in the webpage) to the browser's internal form submission process.
This approach has several key benefits.
First, as it happens outside the web page, no DOM attacker-controlled scripts can interfere with the process.
Second, it will catch all web requests, not just those created through form submission, providing better support if the website submits the values using a client-side script.

To implement this functionality, the \texttt{webRequest} API is used.
While this API does not support web request body modification and likely never will (see \S\ref{sec:webrequest}),
it is possible to achieve the desired functionality at the \texttt{onBeforeRequest} stage.
At this stage, the web request's body has not yet been created.
As such, the browser can be instructed to modify its internal copy of the form data (not DOM-accessible), replacing the password nonce with the real password (if the origin is appropriate).
Finally, the generated web request body will contain the user's password.
This design most closely aligns with the design proposed by Stock and Johns~\cite{stock2014protecting}.

This design effectively adds the ability to modify web request bodies.
Allowing such modifications could introduce new security and performance issues~\cite{chromium91191webrequest,mozilla1376155webrequest}.
Thus, the browser should control the replacement rather than the password manager (i.e., the extension).
Ideally, the extension would list the following items when requesting a replacement:
\begin{enumerate*}[label=\emph{(\roman*)}, itemjoin={{, }}, ]
	\item the password nonce
	\item the replacement password
	\item the origin (scheme, host, port) associated with the password
	\item the name of the field that was autofilled with the password nonce
\end{enumerate*}.

The browser will search through the key-value pairs in the \texttt{FormSubmission} object, making the requested substitution if and only if
\begin{enumerate*}[label=\emph{(\alph*)}, itemjoin={{, }}, ]
	\item the indicated field's value matches the password nonce exactly
	\item the web request will be sent to the indicated origin
\end{enumerate*}.
Requiring (a) may lead to some compatibility issues if a script creates a web request using a different field name, though we didn't encounter this in our testing (see \S\ref{sec:evaluation}).
If necessary, this requirement could be dropped, though it might have unexpected consequences.\footnote{We could not identify any security issues that couldn't already be caused by malicious client-side scripts.}
Additionally, because the replacement happens within the \texttt{FormSubmission} object, proper sanitization and encoding of the key-value pairs will occur, preventing the substitution of one value from being later treated as a different value or multiple values in the request body.

Finally, we note that unlike Design \#3, this design allows copying and pasting of password nonces from the password manager.
As the web request API examines all outgoing web requests, replacement of nonces is still possible.
Still, there will need to be functionality allowing users to copy real passwords if they are to be entered outside of the browser (though this can be made harder to activate, incentivizing the use of nonces where possible).

\paragraph{Security evaluation.}
As the replacement of the password nonce happens after any client-side scripts are allowed to execute and the actual password is never included in the DOM, neither the honest-but-curious entity nor the DOM attacker can exfiltrate the password from the DOM.
The DOM attacker could change the destination to which form data will be sent, but this will result in an origin not associated with the password, preventing replacement.

As with Design \#3, the actual password is still visible in the web request body, allowing exfiltration by an extension attacker (see Figure~\ref{fig:exfiltrate}).
Also, a MitM attacker can steal the password during transmission.
Unlike Design \#3, this design provides strong protection against a phisher. Not only will autofill be prevented but even if a password nonce is copied into the phishing website, it will not be replaced as it will not have the appropriate origin.

\paragraph{Deployability evaluation.}
This design does not require any changes to websites.
It does require changes to the extension API and the form submission code (both executed in the UI processes), but not to code that runs in the browser's core process.
It requires changes to the browser's content code but not its core code.

\subsubsection{Design \#5: Browser-Based Nonce Injection}\label{sec:design5}
To protect against malicious extensions that can read web request bodies, the replacement code must execute after web requests are last allowed to see the web request body.
Our investigation determined that this is best achieved by moving the replacement code into the browser's networking code.
While this has significant implementational challenges (the network code is in a different security domain from where extensions operate), it protects against malicious extensions.

In this design, password managers register autofilled nonces with the browser.
Then, when the browser's networking code is about to send the web request to the operating system for transmission over the network, it first scans for any registered nonces.
If it finds them, it asks the managers for replacement credentials.
At this point, replacement happens as it did in Design \#4, with the manager listing the actual password, the origin, and the field, and the browser only making the changes if all these values match as expected.
Note, even though this replacement happens after the web request body has been formed, modifications still happen through an object, ensuring proper sanitization and encoding occur.

\paragraph{Security evaluation.}
As with Design \#4, and for the same reasons, Design \#5 is impervious from DOM-based password exfiltration (honest-but-curious entity, DOM attacker).
As the password nonce replacement happens after extensions can view the web request body, this design protects against malicious extensions.
It performs the same as Design \#4 and for the same reasons regarding MitM attacks and phishers.

\paragraph{Deployability evaluation.}
This design does not require any changes to websites.
It does require changes to the code that executes in the browser's core process but not to any code executed in the UI processes.
Most standard browsers are built on standard HTML specifications using WebIDL~\cite{webidl}, so deployment into other browsers would be very simple.

\subsection{Discussion}

Examining the five designs, we see that all five designs improve upon the security of the current password entry process used by password managers.
Of these, Design \#1 (zero-knowledge proofs) has the best security.
However, its reliance on websites supporting this technique likely means it is a non-starter.

Looking at the remaining designs, Design \#3 stands out as requiring no changes to websites or the browser.
While it has limited security benefits, it still does better than current practices and is certainly something password managers could explore.

However, we find Design \#5 to be the most compelling.
While it does require modifying the browser, it comes the closest to zero-knowledge proofs in terms of security, theoretically preventing credential exfiltration in all cases except against a MitM attacker who can compromise the security of TLS (a high bar).
As such, this is the design we chose to implement and discuss for the remainder of the paper.

Lastly, we compare these designs against two-factor authentication (2FA) since some may consider these a solution to the problem of password exfiltration.
Two-factor authentication does not do anything to prevent password exfiltration, only limiting the impact of that exfiltration~\cite{bonneau2012quest}.
Research has also shown that many 2FA schemes are vulnerable to the exfiltration of the codes created by the something-you-have factor~\cite{dmitrienko2014security,ulqinaku2021real}.
While there are approaches to secure 2FA against phishing of the secondary factor, this still does not address the issue of password theft.
While it does lessen the impact of a stolen password, it does not remove it, as the stolen password (or a close variant) may be used on other sites.
From a deployability standpoint, 2FA requires more changes than any of our proposed solutions, suggesting they could see faster and more widespread adoption than 2FA.


\section{Implementation}

We implemented Designs \#3--\#5 to demonstrate their feasibility and to perform security evaluations.
In this section, we provide the implementation details of Design \#5, which we found to be the most effective solution.

\begin{figure*}
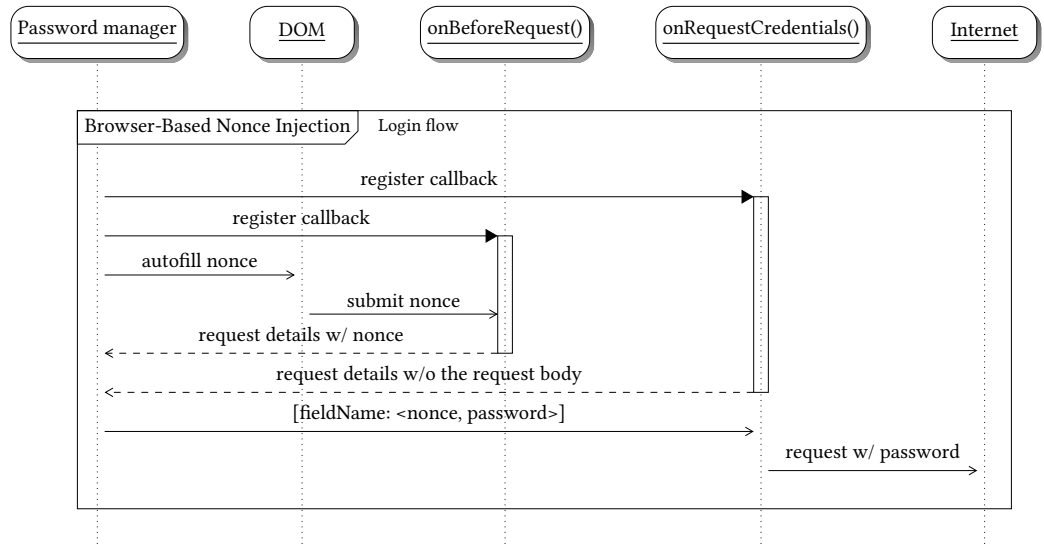

	\centering
	\adjustbox{max width=.8\textwidth}{
		\begin{sequencediagram}
			\tikzset{inststyle/.append style={
					drop shadow={top color=gray, bottom color=white}, 
					rounded corners=2.0ex
				}
			}   
			
			\newinst{p}{Password manager}
			\newinst[1]{d}{DOM}
			
			\newinst[1]{b}{onBeforeRequest()}
			\newinst[1]{a}{onRequestCredentials()}
			\newinst[1]{i}{Internet}
			
			\begin {sdblock}{Browser-Based Nonce Injection}{Login flow}
			\begin{call}{p}{register callback}{a}{request details w/o the request body}
				\begin{call}{p}{register callback}{b}{request details w/ nonce}
					\mess{p}{autofill nonce}{d}
					\mess{d}{submit nonce}{b}
					
				\end{call}
			\end{call}
			\mess{p}{[fieldName: <nonce, password>]}{a}
			\mess{a}{request w/ password}{i}
		\end{sdblock}
	\end{sequencediagram}
}

\caption{This diagram gives the flow for autofilling and replacing nonces as implemented by Design \#5.}
\label{fig:system}
\end{figure*}

To implement Design \#5, we modified Mozilla Firefox 107.0 and the Bitwarden password manager.
We added a write-only \texttt{onRequestCredentials} API to the browser that enables the password manager to detect the submission of an inserted nonce to the server, replacing it with the appropriate credential just before the request is sent over the wire and after other extensions could read the modified request body.
A diagram of the modified flow is given in Figure~\ref{fig:system}.

\subsection{Getting Setup}
First, the password manager will register two callbacks, one for \texttt{onBeforeRequest} and one for the new \texttt{onRequestCredential}.
While we could have collapsed the functionality of both of these callbacks into \texttt{onRequestCredential}, doing so would have required a larger change for how callbacks are handled in the \texttt{webRequest} API.
As this could lead to unintended regressions, we decided on the approach that changed fewer lines of code in the browser, even if it led to a few more lines of code in the password manager implementations.

Next, as needed, the password manager will autofill a nonce in place of a password.
After doing so, it will internally store an association between the web page and 
\begin{enumerate*}[label=(\roman*),itemjoin={,\space{}},itemjoin={,\space{}and\space{}}]
	\item the nonce
	\item the name of the field storing the nonce
	\item the password manager entry that is being autofilled 
\end{enumerate*}

\subsection{\texttt{onBeforeRequest}}
Eventually, the page will be submitted and a \texttt{webRequest} will be created by the browser, with this \texttt{webRequest} containing the autofilled nonce.
This will cause the callback registered with \texttt{onBeforeRequest} to be triggered.
Within this callback, the password manager will be able to see the details of the \texttt{webRequest}, including the destination URL, the HTTP method used, and the request body.
Using this information, the password manager determines if (a) there is a nonce associated with the web page that generated the \texttt{webRequest}, (b) the nonce is in the request body, and (c) replacing the nonce would be safe.
If all these checks pass, the password manager will internally store an association between the \texttt{webRequest} and 
\begin{enumerate*}[label=(\roman*),itemjoin={,\space{}},itemjoin={,\space{}and\space{}}]
	\item the nonce
	\item the name of the field storing the nonce
	\item the password manager entry that is being autofilled 
\end{enumerate*}

Based on recommendations from the research literature~\cite{stock2014protecting,li2014emperor,silver2014password,oesch2020that,oesch2021emperors} and to prevent a reflection attack discussed in \S\ref{sec:reflection}, we recommend that password managers make the following safety checks:

\begin{enumerate}
	\item Check that the web page is not displayed in an \texttt{iFrame}.
	
	\item Check that the submission channel does not use HTTP or an insecure HTTPS connection.
	
	\item Check that the origin (protocol, domain, port) matches the origin identified by the password manager entry being autofilled. Even better, the password manager can store the exact URL where credentials should be submitted, only submitting passwords to that URL.\footnote{Determining the exact URL could be done by having password managers store associations for popular websites, storing prior successful submission locations, or crowdsourcing the creation of associations.}
	
	\item Check that the nonce is not in the GET parameters.
	
	\item Check that the name of the field storing the nonce remains unchanged since the nonce was autofilled there (which itself should have been checked before autofilling~\cite{oesch2020that,oesch2021emperors}).
\end{enumerate}

\subsection{\texttt{onRequestCredential}}
Immediately before sending the \texttt{webRequest} over the wire, the callback registered with \texttt{onRequestCredential} will be triggered.
This callback once again receives request details as input, though in this case the request body has been stripped out.
This is necessary to prevent extensions from reading any nonce substitutions that might have been made by other \texttt{onRequestCredential} callbacks.
In this step, if the password manager has a nonce associated with the current \texttt{webRequest}, it will simply return the associated nonce, password, field storing the nonce, and the URL that the password manager expects the password to be submitted to.

The password manager will then take this information and use it to locate and replace the nonce in the request body.
To prevent unintended (or malicious) consequences from replacing the nonce, the browser will only make this substitution if,
\begin{enumerate}[label=\roman*)]
	\item The field name storing the nonce exactly matches the field name specified by the password manager.
	\item The field value exactly matches the nonce with no additional characters.
	\item The submission URL matches the origin (protocol, domain, port) of the URL provided by the password manager.
\end{enumerate}

While the password manager should have already checked these items before submitting the password to be replaced, we have the browser also check these items to ensure that they will be checked.
This is an important form of defense in depth~\cite{oesch2021emperors}.
If any of these checks fail, no substitution is made, and the request will be sent still containing the nonce.


\section{Evaluations}\label{sec:evaluation}

We evaluated our implementations of Designs \#3, \#4, and \#5 in terms of security.
For Design \#5, we also implemented it in terms of functionality and overhead.

\subsection{Security Evaluation} \label{sec:security-evaluation}

To evaluate the security of our implementations, we first created a basic PHP login page that takes in a username and password as form input and logs the credentials in plaintext for verification purposes.
To test this web page, we would autofill with the password and then submit the web page.

To simulate a DOM-based attack, we injected JavaScript on the test web page that scrapes the password field before page submission.
For this attacker, all three designs prevented the script from accessing the password.
However, Design \#3 only succeeded at stopping this attack because we simulated an attacker who was unaware of the defense.
By simulating an attacker aware of Defense \#3, it is trivial to steal the password.

To simulate a malicious extension, we built and installed a browser extension that logs all outgoing requests using the onBeforeRequest web extension API.
For both Designs \#3 and \#4, the password was successfully stolen.
On the other hand, Design \#5 was successful in preventing password exfiltration, with only the nonces being exfiltrated.
This is true even if we allowed the attack to register a \texttt{onReplaceCredential} callback (which has no access to the request body).

To further verify our results, we conducted these same tests on the login pages for the top 10 Alexa websites.
In each case, our results were the same as our initial test.

\subsubsection{Reflection Attack}
\label{sec:reflection}

An attacker who is aware of Defense \#5 could attempt to conduct a \emph{reflection attack}, wherein they attempt to trick the browser into submitting the real password to a web page that displays those values to the user.
For example, the attacker could change the field name for the password to ``username'' for web pages where an error is displayed when the username doesn't exist (e.g., ``\$\{username\} doesn't exist'').
Alternatively, the attacker can also modify the submission URL to point to a page where it is more likely that request body values will be reflected on the web page.

In each of these cases, the attacker is relying on the password manager and the browser to approve the replacement of the nonce even though the password will be sent to an unsafe location.
To address this attack, our implementation requires the password manager to (i) specify the origin associated with the password and (ii) identify the field storing the nonce.
If either of these values does not match, the nonce will not be replaced with the password.

For even better security, the password manager can store and check the exact URL and field name that should be used to submit the password.
This completely foils this reflection attack.
As several password managers are already checking these items before autofilling passwords~\cite{oesch2020that}, we believe it would be reasonable for them to implement these checks.

\subsection{Functional Evaluation}
\label{sec:functional-evaluation}

To evaluate the functionality of our browser-based nonce injection implementation, we conducted tests on real websites to ensure that our implementation does not break their authentication flows

Using the Alexa top 1000 sites from May 1, 2022, we ran a Selenium script that started at the root page of each domain and traversed all links on the page up to a maximum depth of three, search for login pages.
In total, we identified login pages on 623 sites.
After filtering for the subdomains of the same website, like Google and Amazon for different countries, we were left with 573 unique login pages.

\begin{figure}[h]
	\includegraphics[width=\linewidth]{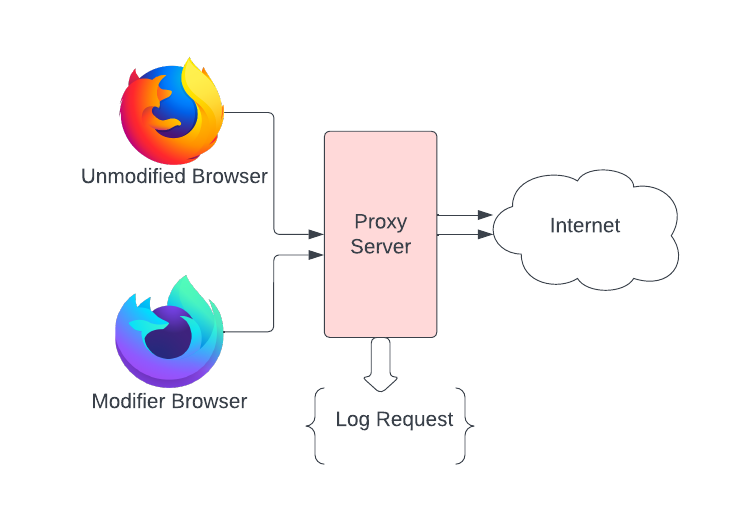}
	\caption{Functional Evaluation Architecture}
	\label{fig:testing_arch}
\end{figure}

For all 573 login pages, we used a Selenium script to submit credentials from both an unmodified Firefox browser and from a modified version of Firefox implementing Design \#5.
As illustrated in Figure~\ref{fig:testing_arch}, we set up a proxy server to record all the outgoing web requests from the browsers and save the request body. 
We then compared the credentials in the authentication requests sent from both browsers.

In 554 of 573 cases (97\%), there was no difference in the credentials submitted to the websites.
11 sites (2\%) generated an integrity check value based on the autofilled nonce.
In these cases, the nonce would still be replaced by the real password, but the integrity check value would no longer match.
While we did not confirm that this would cause the login to fail (we didn't have actual accounts on these sites), we still consider these as failed cases.
Finally, 7 websites (1\%) either hashed or base-64 encoded the nonce, which prevented the password manager from replacing it with the real password.

While 100\% compatibility would be ideal, \emph{being able to improve the security of 97\% is still a significant step forward.}
In practice, password managers could keep a list of websites that don't support secure autofill and not use it for those websites, preventing any functionality regressions.
Moreover, as our testing shows, identifying non-compliant websites can be fully automated, making the creation of such a list highly reasonable.
Additionally, if our solution was adopted, we would hope websites would abandon their ad hoc attempts at secure password entry in favor of the more robust security offered by our secure password autofill system.

\subsection{Overhead Evaluation}
\label{sec:overhead-evaluation}

To measure the overhead incurred by our implementation, we examined the logs for HttpBaseChannel~\cite{httpbasechannel} generated during our functional evaluation of our modified browser.
These logs give exact measurements for each stage of the \texttt{webReuqest} lifecycle, allowing us to pinpoint the processing time used by our code.

On average, the \texttt{webRequest} lifetime was 4.5222 seconds, with our code accounting for 0.443 seconds, representing 10.6\% of the total request duration.
This overhead can entirely be explained by the time needed to destroy and recreate the request stream's body when replacing the nonce with the passwords.
In other words, there is no meaningful overhead when nonce replacement is not needed.
As such, when a password has not been autofilled (the vast majority of cases) there is no meaningful difference in the time taken to submit a \texttt{webRequest}.


\section{Discussion}
\label{sec:discussion}

Below we discuss implications and limitations related to our design and implementation.

\subsection{Deployment and Adoption}
We worked with Mozilla Firefox developers for implementation details on our design ideas.
With the future goal of merging the feature into Mozilla Firefox, we have made careful consideration for changes made in the code to be within acceptable standards for the Firefox repository, getting it reviewed with the developers wherever necessary.
We hope that this will speed deployment of our design.

It is also important to consider that users often reject security advice if they perceive the effort required to implement it outweighs the extra security gained~\cite{herley2009so}.
Our proposed defenses for improving the credential entry workflow do not require any changes to the login user experience, which may increase the likelihood of adopting these protocols.
Additionally, our design only requires small changes to password managers and web browsers, which could be easier to deploy given that both entities have an interest in increasing users' security.

\subsection{Securing Manual Password Entry} 
Protecting manual password entry is challenging, particularly as an attacker can inject client-side scripts (honest-but-curious entity, DOM attacker) that listen and record keystrokes (i.e., a key logger).
One possible way to secure manual password entry would be to allow the browser to enter a special password entry mode using a conditioned-safe ceremony~\cite{karlof2009conditioned}.
In this mode, the browser would record the user's keystrokes, replacing each stroke with one or more random password nonce characters.
The browser would also prevent any scripts or extensions from recording keystrokes while the password entry mode is active.
After the user finishes entering the password, they would leave the password entry mode.

While this approach could protect manually entered passwords, research would be needed to make sure it works in practice.
First, care will need to be taken in selecting the conditioned safe ceremony to ensure that users always enable it when needed.
Second, it will be necessary to design the password entry mode such that it is clear when it is activated~\cite{dhamija2005battle}.
Third, research would be needed to explore how users could be made aware of and encouraged to use this functionality.
Lastly, user studies would be needed to ensure that it has sufficiently high usability to encourage users to make use of this functionality.

\subsection{Securing Credential Entry for Other Authentication Mechanisms}
Even though our study focuses on secure password entry for login forms, other authentication mechanisms also require secure credential entry. For example, research has shown that browser extensions can replace the public keys during FIDO2 registration, allowing attackers to register their tokens to victims' accounts \cite{hu2016security}. A secure credential channel for FIDO2 within the browser could mitigate such attacks. Furthermore, future work could explore additional authentication methods that benefit from a secure credential entry channel. A unified, secure channel for all security-related operations in the browser could have broad applicability and improve the overall security posture of users.

\subsection{Leveraging Nonces for Attack Detection}
In our design, nonces ensure the confidentiality of credentials.
However, we note that it may also be possible to use nonces to detect attacks.
For example, if the browser detects that a nonce is included in a web request body, but that there is no appropriate substitution, this indicates a password exfiltration attack.
The browser could then analyze a webpage's DOM for malicious scripts and potentially share this information with websites.
Similarly, it could search for potentially malicious extensions that are exfiltrating the nonce.
Alternatively, if websites can differentiate between nonces and passwords, this would also allow them to detect that an attack has occurred when they received a nonce.
In either case, the fear of detection may be enough to reduce the likelihood that an attacker will attempt to steal credentials.

In either case, research will be needed to identify how the browser or website can distinguish nonces, while also preventing the attacker from doing so (if the attacker realizes they are stealing a nonce, they won't steal it to avoid detection).
Similarly, research will be needed to ensure that any forensic analysis happens in a way to minimizes risks to user privacy.

\subsection{Denial of Service for Nonce Injection}
An attacker could detect nonces entered onto a web page's DOM and modify them.
This would cause the authentication flow to fail, as the original nonce would no longer exist and the real password would never be substituted.
After repeated failed authentication attempts, a user may assume that there is something wrong with their password manager and manually enter their password, opening their password up to exfiltration.
It is important to note that this is not additional behavior due to our solution as extensions already have the capability to alter DOM input.
Future work could explore this issue to measure user behavior when encountering a DoS attack and explore potential solutions to mitigate its impact.

\subsection{User Confusion}
User confusion is a potential issue with nonce injection defenses.
If users view the autofilled password, they may realize it is not their password, especially if they don't use a generated password.
This could lead to confusion.
Although this issue may not be that prevalent because users are less likely to investigate passwords inserted through a password manager, it is still prudent to study the behavior of users when they encounter such confusion.

While the shadow DOM is not sufficiently secure to be used to show users their actual password~\cite{ruoti2016messageguard}, the browser could add a similar utility to securely display the real password to the user, without allowing that password to be accessed by malicious client-side scripts or extensions.
Future research could also investigate this potential solution.


\section{Conclusion}

There are many avenues for adversaries to exfiltrate passwords: through client-side scripts, the \texttt{webRequest} API, during network transmission, and through phishing.
In this paper, we explore how the password manager can take an active role in protecting passwords from theft.
To this end, we identify a strong threat model for password exfiltration.
We recommend that future work on this topic also consider a threat model at least this strong.

Based on this threat model we conduct a design space exploration, and identify five potential methods that password managers can use to secure password entry.
Of these, the most secure design involves having password managers authenticate directly with websites using a zero-knowledge proof.
While we do not pursue this implementation further in this paper, we still think it is compelling, and recommend that password managers work with websites to further explore this design.

Instead, we settle on a design based on password nonce injection.
This design does not require modifying websites but still provides most of the security benefits of using a zero-knowledge proof.
Critically, this design requires no change in user behavior or awareness.
We believe that both of these properties are necessary to promote the possibility of adoption.

To verify that this design is feasible, we implemented it in Firefox and BitWarden.
We then conduct security and functionality evaluations of our prototypes, demonstrating that they can stop credential exfiltration by malicious client-side scripts and browser extensions.
We also demonstrate that it works with 97\% of the Alexa top 1000 websites.

While not perfect, our implementation is working, publicly available code that can already improve the security of autofill on the vast majority of websites.
For the remaining websites, it is easy to automatically detect compatibility issues and not use our secure autofill API, preventing any functionality regressions.
As such, our work not only pushes forward our scientific understanding of this area but makes an important practical contribution.



\bibliographystyle{ACM-Reference-Format}
\bibliography{latex/bibtex/authentication,latex/bibtex/publications,paper}

%
%

\end{document}